\documentstyle[multicol,aps,epsf,here]{revtex} 
\tighten
\newcommand{\PostScript}[7]{
\begin{figure}[H]
\begin{center} 
\leavevmode
\epsfysize=#1cm
\vspace{#2cm}
\epsfbox{#3}
\par
\parbox{#5cm}{
\vspace{#4cm}
\caption[figure]{\renewcommand{\baselinestretch}{1} \small \normalsize #6}
\label{#7}}
\end{center}
\end{figure}
}

\begin{document}

\begin{multicols}{2}
\noindent
{\Large\bf\sf Rotational and vibrational spectra of quantum rings}

\medskip
\noindent
{\bf M. Koskinen, M. Manninen, B. Mottelson$^*$ and S.M. Reimann}\\      
\noindent
{\small \it Department of Physics, University of Jyv\"askyl\"a,\\
40351 Jyv\"askyl\"a, Finland}\\
{\small \it $^*$NORDITA, Blegdamsvej 17, 2100 Copenhagen, Denmark}

\medskip

\narrowtext
{\bf One can confine the two-dimensional electron gas 
in semiconductor heterostructures electrostatically or by etching techniques 
such that a small electron island is formed.
These man-made ``artificial atoms'' provide the experimental 
realization of a text-book example of many-particle physics: a finite 
number of quantum particles in a trap. 
Much effort was spent on making such "quantum dots" smaller 
and going from the mesoscopic to the quantum 
regime~\cite{tarucha,kouwenhoven}.
Far-reaching analogies to the physics of atoms, nuclei or metal clusters were 
obvious from the very beginning: The concepts of shell structure and Hund's 
rules were found~\cite{tarucha} to apply -- just as in real atoms!
In this Letter, we report the discovery that electrons confined 
in ring-shaped quantum dots form rather rigid molecules with antiferromagnetic 
order in the ground state. This can be seen best from an analysis 
of the rotational and vibrational excitations.}

While the independent-particle picture was successful
in describing the electronic structure for rather large particle densities,
for more dilute systems or in stronger magnetic fields correlation
effects are of crucial importance.
Configuration-interaction (CI) calculations, which have a long tradition in 
quantum chemistry and cluster physics, were then much used~\cite{exact}.  
Although these so-called ``exact'' calculations are numerically demanding 
and limited to the smallest sizes, they still are able to provide 
significant insight into the many-body phenomena that occur 
in these finite fermion systems with reduced dimensionality.
In this Letter we apply CI techniques to investigate the electronic structure
of {\it quantum rings} that contain up to seven electrons. 
Usually the confinement of small, two-dimensional quantum dots 
is to a very good approximation harmonic. Correspondingly, we model 
quantum rings as they are realized in the laboratory 
by a potential of the form $V(r) = {1\over 2} m^* \omega _0^2 (r-r_0)^2$.
For moderate confinement this potential
corresponds to an harmonic dot with its center removed.
Ground and excited states of $N$ electrons trapped in the potential 
$V(r)$ are determined from numerical diagonalization 
as a function of the total angular momentum. 
Surprisingly, at electron densities and strengths of the ring confinement 
where one should expect electron {\it liquid} behavior, a model which 
assumes {\it localization} of the electrons in the ring
is successful in analyzing the many-body spectra.
Group-theoretical methods familiar from molecular physics provide the 
necessary tools to uncover rotational and vibrational 
structures in the spectra. 
The spin sequence and energies of the low-lying states for given angular 
momentum can be understood from the symmetry associated with the electronic 
ground state configuration. It is intriguing that the success 
of the simple rigid-rotor model for the low-lying states 
is {\it not} limited to a regime where the system becomes very 
one-dimensional. 
The fact that the electrons behave as if they were localized in the ring 
is also reflected in a remarkable agreement of the CI results with 
the Hubbard or Heisenberg model. Localization at large electron densities 
has earlier been discussed in parabolic quantum dots, where the interpretation
is not yet conclusive~\cite{wigner}. 
We write for the Hamiltonian 
\begin{equation}
H=\sum _{i=1}^N \biggl [ -{\hbar ^2\over 2m^*} \nabla _i^2 + V(r_i)
\biggr ] + \sum _{i<j}^N {e^2\over 4\pi \varepsilon _0 \varepsilon} 
{1\over \mid {\bf r}_i - {\bf r}_j\mid }~,\nonumber
\end{equation}
where $m^*$ and $\varepsilon $ are the effective mass and the 
dielectric constant of the corresponding semiconductor material.
The parameters that determine the properties of the 
quantum ring are the number of electrons $N$, the radius of the 
ring $r_0$ and the strength $w_0$ of the harmonic confinement in 
the radial direction. 
The quantities $r_0$ and $\omega _0$ are related 
to the more fundamental quantities $r_s$, 
the one-dimensional density parameter which describes the 
particle density $n = 1/(2r_s)$ along the ring (thus $r_0=Nr_s/\pi $)
and $C_F$, a dimensionless parameter that measures the degree of
one-dimensionality. $C_F$ essentially describes the excitation energy 
of the next radial mode $\hbar \omega _0$, which is defined to be 
$C_F$ times the (1D) Fermi energy. We thus obtain
$\hbar\omega _0 = C_F \hbar^2 \pi ^2/(32 m^* r_s^2 )~.$
The higher the value of $C_F$, the more the radial modes are  
frozen in their ground states. Thus, the ring is narrower 
for larger $C_F$. For the CI calculation,
the spatial single-particle states of the Fock space are chosen
to be eigenstates of the single-particle part of the Hamiltonian $H$. 
We expand them in the harmonic oscillator basis.
According to their eigenenergies, from 30 to about 50 lowest 
single-particle states
are selected to span the Fock space. Typically this means 
that for lower angular momentum states several radial quantum numbers
$n=0,1,2,3$ are included, whereas the higher angular momentum states 
$l=\pm 6,\dots,\pm 10$ have only $n=0$.
To set up the Fock states for diagonalization, we sample over the full space
with a fixed number of spin down and spin up electrons, 
$N_{\downarrow} + N_{\uparrow } = N$~.
From this sampling, only those states with a given total 
orbital angular momentum 
and a configuration energy (corresponding to the sum of occupied 
single-particle energies) less than the specified cutoff energy $e_c$ 
are selected. The purpose was to choose only the most important 
Fock states from the full basis and hereby reducing the matrix dimension
to a size $d\stackrel {<}{\sim} 2\cdot10^5$.
To obtain all the eigenstates with different total spin,
we have to set $N_{\downarrow} = N_{\uparrow} =N/2$ for even particle 
numbers ($S_z=0$, all states with different total spin have this 
component), and analogously $N_{\downarrow } = N_{\uparrow }\pm 1$
for odd numbers.
Once the active Fock states have been specified, the Hamiltonian matrix is 
calculated. For diagonalization we use the Arpack library~\cite{arpack} 
suitable for large, sparse matrices. Finally, the total spin of
each eigenvector is determined by calculating the expectation value
of the $\hat S^2$ operator. 
The many-particle states are characterized by the total orbital angular 
momentum $M$ and the total spin $S$. 
The lowest energy at given angular momentum $M$ defines 
the so-called ``yrast'' line, a terminology that was introduced in nuclear
physics many years ago and comes from the Swedish word for "the most dizzy".

To begin with, let us look at the energy spectra of a ring with 
six electrons at a density corresponding to $r_s=2 a_B^*$.
(Throughout this paper, we use effective atomic units.
Taking GaAs as an example, the units of length and energy
are $a_B^*=9.8$nm and a.u.$^*=12$meV.)
For $C_F=4$, the strength of the ring confinement is moderate, 
as it can be seen from the density distribution displayed in
Fig.~\ref{f1}.
\PostScript{5}{0}{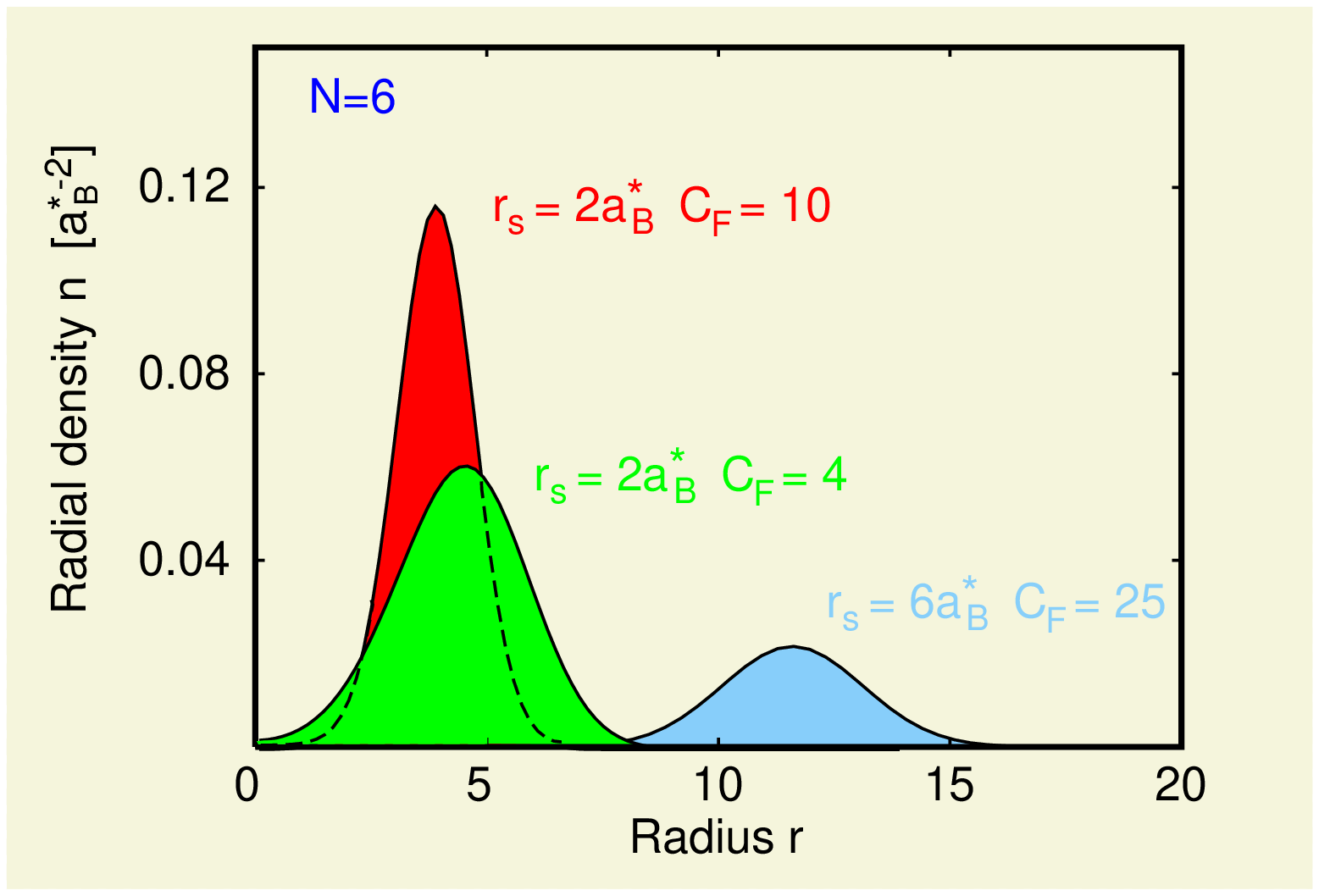}{0.5}{14}{\small\sf Density along the 
radial direction for $N=6$ and different parameters 
$r_s$ and $C_F$. The confinement is moderate for $r_s=2$ and $C_F=4$ 
{(\it green)}. The degree of one-dimensionality is enhanced for  
$C_F=10$ {\it (red)}. The low-density regime is displayed for $r_s=6$ 
and $C_F=25$ {\it (blue)}}{f1} 
The system parameters $N$, $r_s$ and $C_F$ in this 
case are chosen such that the rings resemble quite closely structures as they 
can nowadays be made in the laboratory. 
For the six electron ring, Fig.~\ref{f2} shows the 50 lowest 
states for all angular momenta from $M=0$ up to $M=6$. 
The spin configurations are given for the low-lying states.
The ground state for $M=0$ has spin $S=0$ and is followed by a state
with $S=2$, then $S=1$ and, close in energy, another $S=0$ state.
A large energy gap separates these low-lying states from higher bands 
with a much increased density of states. 
We note that the sequence of spins of the states below the gap for $M=0$  
is {\it repeated} at $M=6$. (The energy difference between the two $S=0$ 
states at $M=6$ is slightly reduced, as in a not very narrow ring rotation 
expands the ring.)
Inspecting the grouping of the states below the gap for each of the 
different $M$-values more closely, we see that in a similar way, 
the states for $M=1$ follow the same sequence as those for $M=5$, and 
the $M=2$ spectrum is repeated at $M=4$. 
\PostScript{6}{0}{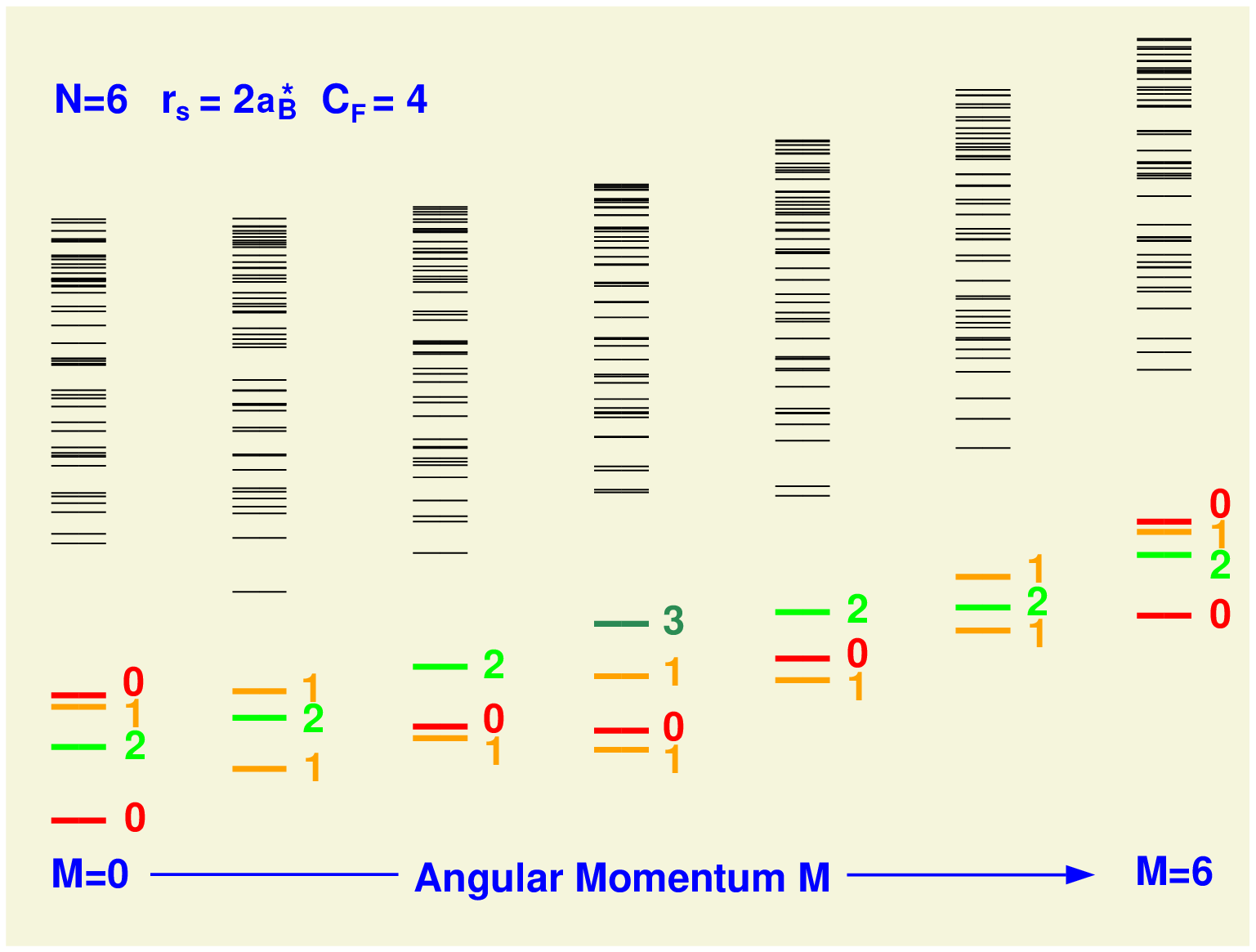}{0.5}{14}{\small\sf Many-body spectra of a 
quantum ring with $N=6$ electrons ($r_s=2a_B^*$ and $C_F=4$). The energy 
difference between the yrast states for $M=0$ and $M=6$ is 0.143~a.u.$^*$. 
The spin $S$ is given for the low-lying states 
(as also indicated by different colors of the levels)}
{f2}
\PostScript{6}{0}{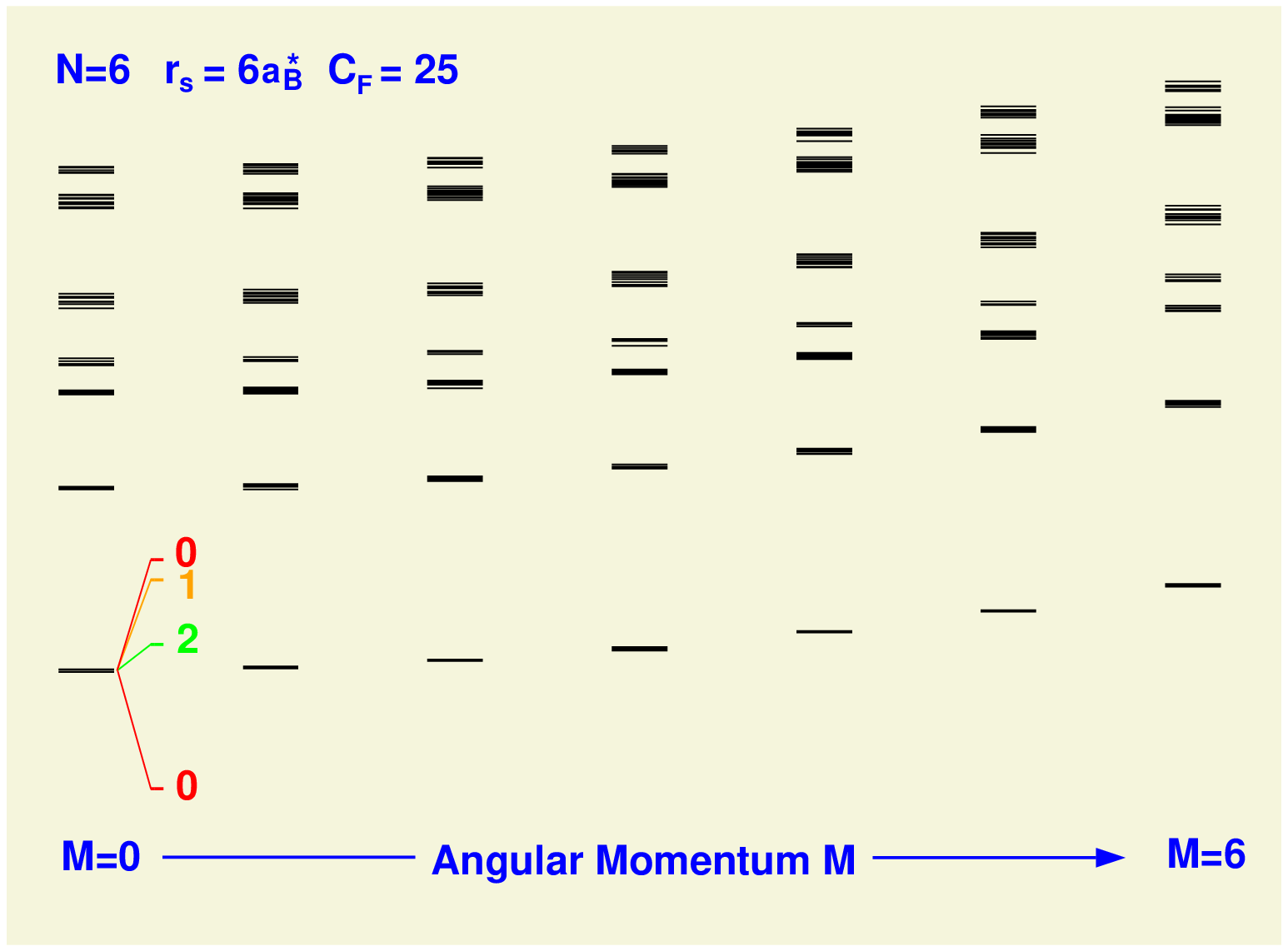}{0.5}{14}{\small\sf 
Spectra of a narrow low-density quantum ring
with six electrons ($r_s=6$ and $C_F=25$), showing the seven lowest
vibrational bands. For the vibrational ground state we show the $M=0$
levels in an magnified scale to demonstrate
that the energy ratios are the same as in the much wider 
rings shown in Figs. 2 and 5.
}{f3}
These sequences of states clearly reveal the spectrum of a 
dynamical system that can be described, at least approximately, in 
terms of independent rotation,
vibration and intrinsic spin degrees of freedom. 
This becomes particularly clear when 
repeating the above calculation for a much lower particle density in a 
more narrow ring. 
We show in Fig.~\ref{f3} the 6-particle spectra for $r_s=6a_B^*$
and $C_F$=25. (The particle density for these parameters is displayed 
in Fig.~\ref{f1}). The spectra now indeed consist of narrow bands.
The lowest band is the vibrational ground state and consist of the 
rotational levels. The different spin levels become almost degenerate, 
as localization has reached a degree where spin-spin interactions 
become less important. Nevertheless, the sequence of spins and their 
relative energetic order is similar to the higher density results shown 
in Fig.~(1), as a plot of the low-lying levels at an enlarged energy scale 
for $M=0$ demonstrates. 
In the following we analyze
the above spectra in terms of the standard effective 
hamiltonian employed in the
interpretation of the rotational and vibrational spectra of planar polygonal
molecules composed of $N$ identical spin $1/2$ fermionic 
atoms~\cite{herzberg}; the
standard description is very slightly modified in order enforce the planarity
and the one-dimensionality of the molecule considered here:
\begin{equation}
H_{eff} = AM^2 +\sum_{a} \hbar\omega_a n_a +  J \sum_{i,j}^{N}
{\bf S}_i \cdot {\bf S}_j  
\label{X}
\end{equation}
where $A$ is the rigid moment of inertia of the $N$ "atoms" located at the $N$
vertices of an equilateral polygon, $ {\omega}_a$ are the frequencies and
$n_a$ $(=0,1,2,\dots)$ the number of excitation quanta of the different normal
modes of vibrations; for odd $N$ there are $\frac{1}{2}(N-1)$ two fold
degenerate normal modes, while for even $N$ there are $\frac{1}{2}N-1$
twofold degenerate  and one non-degenerate  modes. The last term in 
Eq.~(\ref{X}) describes the nearest neighbor spin-spin interaction 
that is a result of the exchange term in the interaction of the 
neighboring fermions. In the expression~(\ref{X})
we have ignored the contribution of a possible spin orbit term.

For the $N=6$ molecule discussed in Figs.~\ref{f2} and~\ref{f3}, 
the $C_{6v}$ symmetry classification of the normal modes 
is (in order of increasing vibrational
frequency) $E_1$ (2-fold degenerate), $E_2$ (2 fold degenerate) and $B_1$
(non-degenerate). The rotational states carry symmetry for 
$M\equiv 0 ~(A_1$ or $A_2$),
$M\equiv 1$ or $5 ~(E_1)$  
$M\equiv 2$ or $4 ~(E_2)$,
$M\equiv 3 ~(B_1$ or $B_2)$, 
where the congruent sign in all these terms
refers to congruent (mod 6), and the total wave function must have $C_{6v}$
symmetry $B_2$ in order to fulfill the requirement of anti-symmetry between all
the electrons.

As we learned from the above analysis, the lowest 
band is of rotational nature, while the bands above the energy gap
represent vibration in addition to rotation.
The three  vibrational modes of such a system have energy ratios 
$\omega_1:\omega_2:\omega_3=1:\sqrt{7/3}:\sqrt{3}$,
in excellent agreement with the three lowest excitations for $M=0$
shown in Fig.~\ref{f3}. The three highest states have then
energies $2\omega_1$, $\omega_1+\omega_2$ and 
$\omega_1+\omega_3$. 

A generic model for localized electrons 
is the Hubbard model~\cite{vollhardt,jefferson} which can be solved exactly 
for a small number of lattice points.
The real electrons in the ring interact with the 
long-range repulsive Coulomb force. Classically the electrons
would form a ring of equidistantly localized electrons.
In the Hubbard model this case corresponds to the so-called
half-filled model, where the number of electrons equals
the number of lattice points.
In the strong interaction limit it
can be transformed to the anti-ferromagnetic Heisenberg model\cite{vollhardt}
leading to the last term of Eq.~(\ref{X}).
\PostScript{9}{0}{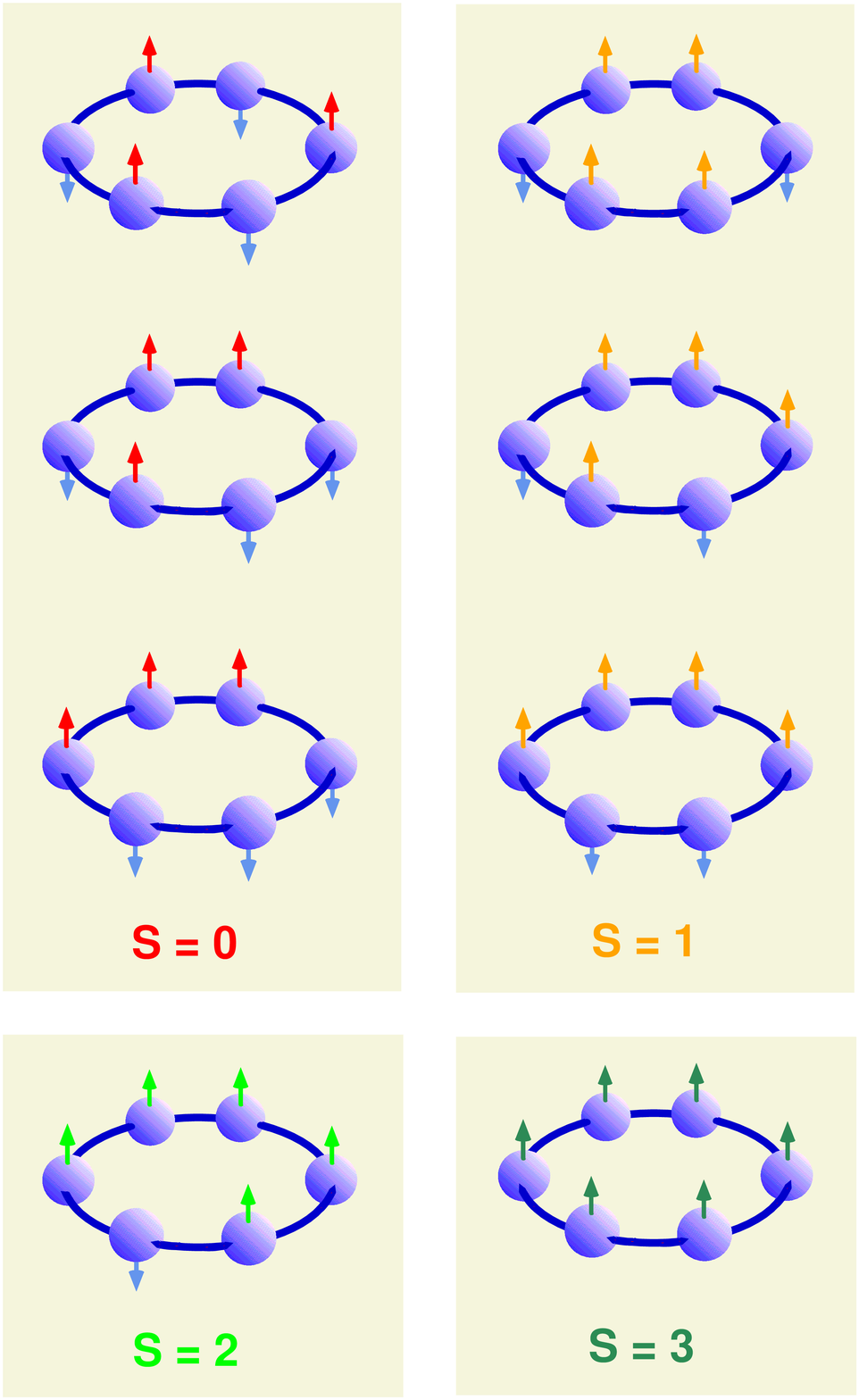}{0.5}{14}{\small\sf Schematic view of the 
spin configurations in a six-electron ring for different values of $S$.}{f4}
On the basis of the results from exact diagonalizations and their 
group-theoretical analysis described above it is now interesting 
to see how well such simple models of localized electrons can describe 
the many-body spectra in these finite, not truly one-dimensional quantum rings.
Fig.~\ref{f5} compares the energy spectra of the 
Heisenberg model to the CI spectra described above.
In the latter case the center-of-mass motion associated to the
orbital angular momentum ($AM^2$-term) is subtracted from 
each level. The coefficient $A$ is determined to make the 
lowest energy eigenvalue corresponding to $M=N$ equal to
that of $M=0$. 
The resulting values of $A$ are very close to 
$A=\pi ^2/(2N^3r_s^2)$ obtained for strictly localized electrons.
The scale of the Heisenberg result is determined
to get the same band width as in the case of real electrons.
Fig.~\ref{f5} displays for electron numbers $N=5, 6$ and 7 at  
$r_s=2a_B^*$ and $C_F=10$ the spectra of the states below the first gap 
which (as analyzed above) all are of rotational nature.  For $N=7$,
corresponding to the larger particle number, there are now more states 
as we have more possibilities for the different spin configurations.
The above comparison shows that indeed, the electrons in a ring
can be rather accurately described as localized electrons:
{\it The whole rotational spectrum close to the yrast line
can be determined by a spin model combined with a rigid
center-of-mass rotation.}

It should be noted that the results shown in Fig.~\ref{f5}
are calculated for a ring with a finite width. Making the ring 
narrower the small disagreement between the 
Heisenberg model and exact calculation vanishes.
We finally mention that the pair correlation function does indicate 
anti-ferromagnetic coupling of the electrons in the ground state. 
(See configuration schematically shown in the top left corner of
Fig.~\ref{f4}). However, the correlation is not as clear as expected from
the rotational structure in the energy spectrum. In the limit of an 
infinitely long one-dimensional Heisenberg model the spin-spin correlation 
decreases as $1/r_{ij}$. 
Consequently, even in the more narrow rings, where the rotational 
and vibrational spectra show a clear localization of electrons, 
the pair correlation function shows only rather weak spin-spin 
correlation.
\PostScript{10}{0}{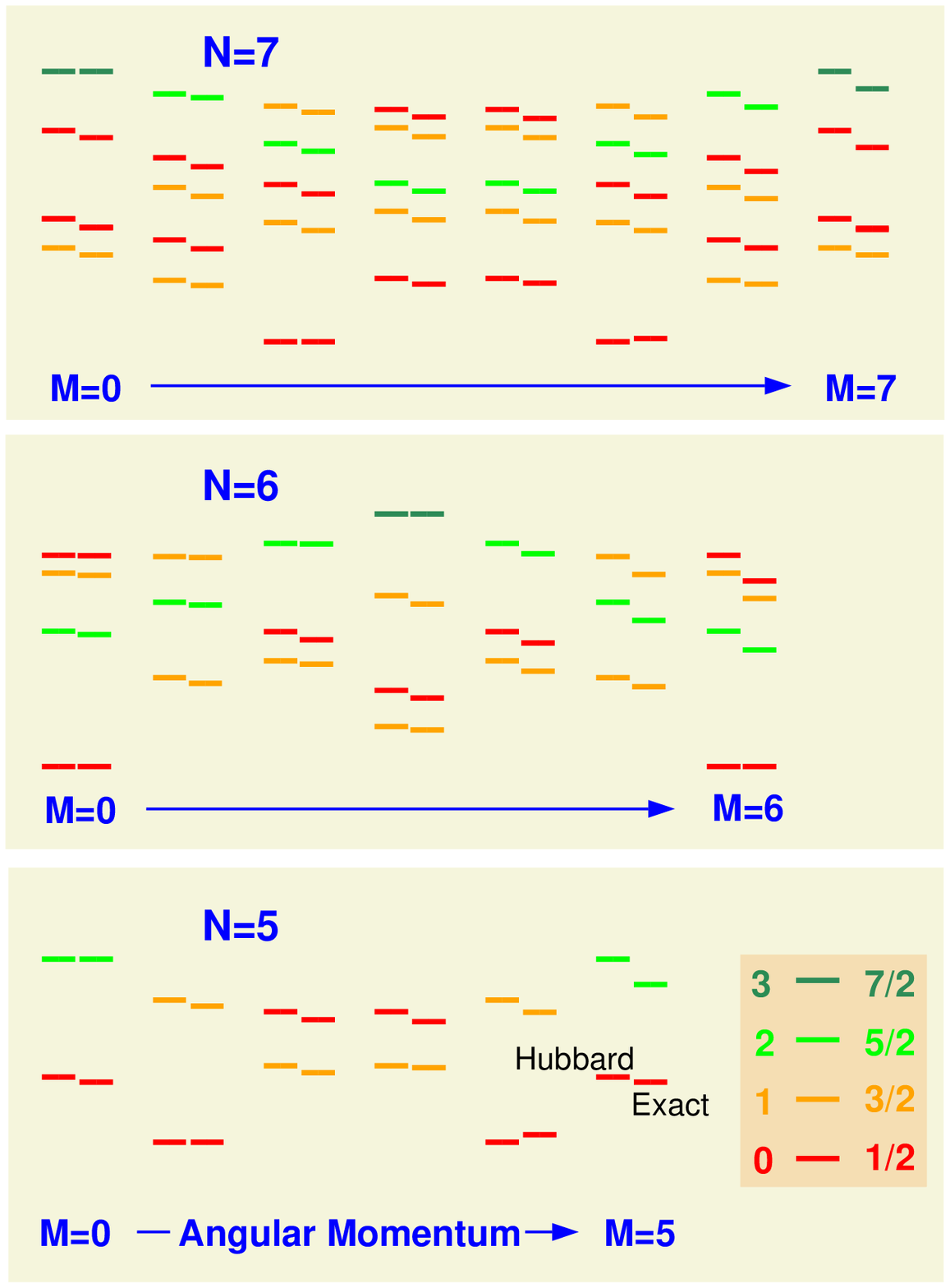}{0.5}{14}{\small\sf Low-lying (rotational) spectra 
of quantum rings with $N=5, 6$ and 7 electrons: Comparsion between 
the Heisenberg model {\it (left columns in each of the spectra for given $M$)} 
and exact diagonalization {\it (right)}. The spins of the 
levels are marked by different colors}{f5}
In conclusion, we reported rotational and vibrational many-body 
spectra of quantum rings confining up to seven electrons. We saw that 
the results obtained by configuration interaction calculations are best
described by assuming localization 
in the internal structure of the ground-state many-body 
wave function. For even 
electron numbers, antiferromagnetic ordering was found for the 
ground state.
Such so-called spin density waves in quantum rings actually were predicted 
by density functional theory~\cite{sdwprl}. 
The many-body spectra of the reported exact diagonalization 
calculations, their analysis by group-theoretical methods and further
the comparison to the Heisenberg model confirm 
that indeed the simple mean-field picture (as it is provided by density 
functional theory) can to a rather large degree correctly map 
out the {\it internal} symmetry of the many-body wave function.
 
This work was supported by the Academy of Finland,
the TMR program  of the European 
Community under contract ERBFMBICT972405 and a NORDITA Nordic Project on 
``Confined Quantum Systems''.

\end{multicols}
\end{document}